# Line-shape study of CO perturbed by N$_2$ with mid-infrared frequency comb-based Fourier-transform spectroscopy

AKIKO NISHIYAMA[1,2,*], GRZEGORZ KOWZAN[1], DOMINIK CHARCZUN[1], ROMAN CIURYŁO[1], NICOLA COLUCCELLI[3,4], PIOTR MASŁOWSKI[1]

1. Institute of Physics, Faculty of Physics, Astronomy and Informatics, Nicolaus Copernicus University in Toruń, Grudziądzka 5, 87-100 Toruń, Poland
2. National Metrology Institute of Japan (NMIJ), National Institute of Advanced Industrial Science and Technology, 1-1-1 Umezono, 305-8563, Tsukuba, Ibaraki, Japan
3. Dipartimento di Fisica - Politecnico di Milano, Piazza Leonardo da Vinci 32, 20133 Milano, Italy
4. Istituto di Fotonica e Nanotecnologie - CNR, Piazza Leonardo da Vinci 32, 20133 Milano, Italy

*Corresponding author: akiko@fizyka.umk.pl

Abstract: We developed a mid-infrared optical frequency comb-based Fourier-transform spectrometer and performed a line-shape study of the fundamental vibrational band of CO perturbed by N$_2$, which is crucial for atmospheric science and astronomical observations. The comb-based FTS enabled us to measure the whole vibrational band with high resolution and precision at several pressures between 10 and 400 Torr. Observed absorption profiles were fitted with the speed-dependent Voigt profile. Collisional broadening, speed-dependent collisional width and shift coefficients are derived. The reliability of our results is established from considerations of systematic errors and comparison with previous studies.

## 1. Introduction

The mid-infrared (mid-IR) wavelength region contains strongly absorbing fundamental vibrational bands of various molecules. It is an important wavelength region in spectroscopic experiments for atmospheric observation, space observation and fundamental physical chemistry. Mid-IR spectroscopy experiments have often relied on Fourier transform infrared (FTIR) spectroscopy to provide broadband molecular spectra. However, due to the intrinsic limitations on spectral resolution and precision of FTIR, the accuracy of spectral data provided by this method is limited. Precise infrared spectroscopy using cw lasers (obtained directly in the mid-IR or converted to that range with difference frequency generation) has also been performed, but the inability of broadband spectral acquisition in reasonable measurement time is the downside of tunable laser-based spectrometers. For these reasons, it has been difficult to provide highly precise spectral data over the entire ro-vibrational bands. The solution to all these problems is optical frequency comb-based spectroscopy.

Over the last 15 years there has been active research into spectroscopic techniques using optical frequency combs as light sources. One such technique is optical frequency comb-based Fourier transform spectroscopy (OFC-FTS), which allows broadband spectral acquisition with high frequency resolution far exceeding that of conventional FTIR [1,2]. The resolution of FTIR is limited to the reciprocal of the finite length of the interferogram, explained as the Fourier transform limit, but the resolution of OFC-FTS is limited only by the mode linewidth of the frequency comb [3,4]. The OFC-FTS apparatus simply replaces the thermal light source with a frequency comb in conventional FTIR that has long been used in spectroscopy laboratories; low installation cost is one of the advantages of OFC-FTS. Mid-IR frequency comb sources are relatively readily available these days. Several types of such sources have been developed, such as quantum cascade laser-based frequency combs oscillating directly at infrared wavelengths [5,6], or frequency-converted near-infrared frequency combs by optical parametric oscillator (OPO) [7,8] or difference frequency generation (DFG) [9-11]. The OFC-FTS is a powerful alternative to FTIR and tunable laser-based spectrometers for broadband high-resolution mid-IR spectroscopy.

Carbon monoxide (CO) has its fundamental vibrational (0-1) band at 4.6 μm and plays an important role in atmospheric physics. It is used as an indicator of abundance of methane, which is important contributor to global warming [12,13]. Further interest in CO is due to its important role as a tracer in various systems such as fossil fuel and biomass combustion processes [14,15] or the atmospheres of satellites in the solar system [16,17]. Therefore, high-precision spectral data of CO is often required for the analysis of atmospheric absorption spectra and astronomical observations. In particular, the spectral data for the collisional line shift and broadening caused by nitrogen molecules ($N_2$), a major component of the atmosphere, is important to the spectroscopy community. To date, spectral data of the 0-1 band of CO-$N_2$ has been provided by FTIR [18-20] and tunable cw-laser measurements [21-24]. Since the CO collisional shifts are approximately ten times smaller than the broadenings, accurate measurement of the pressure shift constants is difficult and has only been reported in a few previous papers [20,24].

In this paper, we present precise collisional line-shape measurements of the 0-1 band of CO-$N_2$ using OFC-FTS. The line shapes of the P1 to P20 and R0 to R20 lines are fitted using speed-dependent Voigt profiles. Collisional shift and broadening coefficients are provided for each line as well as speed-dependent coefficients. For the first time, we provide CO-$N_2$ data compatible with the new standard of spectroscopic data in HITRAN. [25,26]. This standard incorporates the Hartmann-Tran profile [27], which is the recommended form [28] of the speed-dependent hard collision profile [29,30], allowing simultaneous inclusion of speed-dependent effects [31] and the Dicke narrowing [32-34].

## 2. Experiment

### 2.1. Spectroscopic system

The setup of our OFC-FTS (shown in Fig. 1(a)) was also briefly described in conference communication [35] together with initial results and simplified data analysis. We used a home-built mid-IR frequency comb source generated by an optical parametric oscillator (OPO) [8] synchronously pumped by a mode-locked Yb-fiber laser (Yb-OFC) [36] at 1.06 μm, with the repetition frequency ($f_{rep}$) of 125 MHz. The output of the Yb-OFC is amplified in a pre-amplifier with a single-mode Yb-doped fiber (amp1) and a chirped-pulse amplifier (amp2) with an Yb-doped large-mode area double-clad fiber and 10 W pump diode lasers [37]. Afterwards, the pulses are compressed with a transmission grating pair. The input pulse train to the OPO cavity has an average power of 2.7 W and a pulse duration of 175 fs at full-width at half maximum (FWHM). The OPO cavity contains a 5 mm-long MgO-doped periodically poled lithium niobate (PPLN) crystal that provides idler and signal tunable from 2.6 to 4.9 μm and from 1.75 to 1.35 μm, respectively, under quasi-phase matching conditions. The OPO cavity has a free spectral range of 125 MHz and consists of five mirrors with high reflectivity at the wavelength range of the OPO signal. Figure 1 (b) shows idler spectra and corresponding power obtained for different poling periods of the PPLN crystal. Absorption features in the spectra are due to water (from 2.5 to 3.3 μm) and carbon dioxide (at 4.2 μm) in the atmosphere. The output power reaches more than 100 mW for most emission wavelengths including our target wavelength of 4.6 um, which is more than sufficient for absorption spectroscopy.

To stabilize the mode frequencies of the mid-IR frequency comb source, the $f_{rep}$ of the pump laser and the offset frequency ($f_{ceo}$) of the idler are phase-locked to RF reference signals. The idler $f_{ceo}$ is detected as the beat note between the pump laser comb broadened by a photonic crystal fiber (PCF) and the sum frequency (parasitic) of the pump and idler derived at the OPO output by spectral filtering. The phase difference signal is fed back to the OPO cavity length; the cavity length stabilization also avoids drifts in the output spectrum of the mid-IR OFC. Figure 1(c) shows the stabilized $f_{rep}$ and $f_{ceo}$ of the mid-IR OFC as measured with a frequency counter using a gate time of 1 s. Taking into account the residual fluctuations of $f_{rep}$ and $f_{ceo}$, the center frequency of the comb mode at 4.6 um with mode number $n = 520000$ has an overall fluctuation $\delta v_n = 431$ Hz, assuming uncorrelated $f_{rep}$ and $f_{ceo}$ noise. The mid-IR OFC output at about 4.6 μm is first passed through a 72-cm long cell filled with a CO-$N_2$ mixture and then coupled into a home-built FTS system *via* a mid-IR fiber. The FTS system simultaneously records mid-IR OFC and stabilized He-Ne laser interferograms, with the latter used as the path length reference. The interferogram length is set to $c/f_{rep} = 2.4$ m to eliminate the influence of the instrument line shape (ILS) and to achieve high resolution measurements [1] beyond the Fourier limit. The acquisition time of an interferogram is only 11 s, but we average many of them to achieve sufficient signal-to-noise ratio (SNR) for line-shape analysis beyond the Voigt profile. We investigated the scaling of SNR with the number of interferograms (Fig. 2(a)). Based on this, we set our averaging at 50 interferograms, corresponding to about 9 minutes, which coincides with saturation of the SNR level. Longer averaging does not improve the SNR because of the drift of the OPO output spectral envelope. After the measurements of the CO-$N_2$ mixture, the transmission spectra of an empty cell were measured in the same manner to use for background normalization.

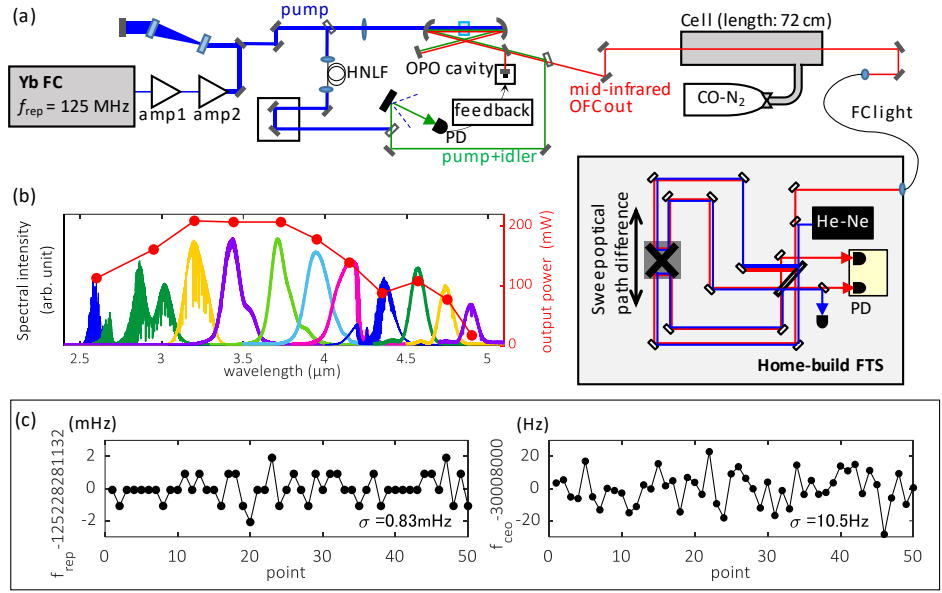

Fig. 1. (a) Experimental setup consisting of a mid-IR comb source and a home-built FTS. amp1: preamplifier, amp2: chirped-pulse amplifier, PCF: photonic crystal fiber, PD: photodiode. (b) Idler output spectra with different PPLN poling periods, and idler output power at each wavelength (red, right axis). (c) Frequency stability of $f_{rep}$ (left) and $f_{ceo}$ (right). The gate time of the frequency counters was 1 s.

Line-shape measurements were performed at CO-$N_2$ gas pressures of 1.3 kPa, 4.0 kPa, 13 kPa, 33 kPa, and 53 kPa (10 Torr, 30 Torr, 100 Torr, 250 Torr, and 400 Torr). We used CO-$N_2$ gas mixtures with an adjusted CO concentration of 1000 ppm for 10 Torr and 300 ppm for 400 Torr measurements so that the transmittance of the strongest absorption feature was approximately 20 % at each pressure. Under all conditions, the CO concentration is low, and the effect of self collisions of CO is negligible. The temperature of the sample gas was measured by a thermistor attached to the cell. During all measurements, the average temperature was 296.8 K with variation of ±0.5 K. The normalized spectra of the CO 0-1 band measured at a CO-$N_2$ mixture pressure of 100 Torr are shown in Fig. 2(b). The SNR was about 1000. Since the spectral range of the instantaneous OPO output is limited by the phase-matching condition of the PPLN crystal, the P and R branches were measured separately. At the pressure of 100 Torr, the half width at half maximum (HWHM) of the absorption lines is about 300 MHz, hence the sampling density of $f_{rep}$ (=125 MHz) is insufficient to accurately probe the line shapes. Therefore, at low pressures we performed measurements at multiple $f_{rep}$ values and interleaved them to obtain sampling densities of 62.5 MHz at 100 Torr, 31.3 MHz at 30 Torr, and 25 MHz at 10 Torr. Interleaving was performed after averaging 50 spectra at each $f_{rep}$ value.

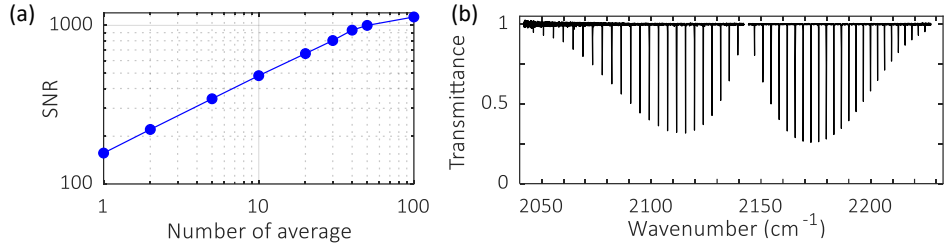

Fig. 2. (a) Improvement of the SNR by averaging. (b) Observed spectra of the 0-1 band of CO. The measurement was performed with 100 Torr of sample gas, averaged 50 times and normalized to reference spectra taken with an empty cell.

## 2.2. Data processing procedure

We performed data processing of the OFC-FTS data according to the procedure described in ref.[2]. The acquired interferograms were cut to the length of $c/f_{rep}$ referenced to the He-Ne laser wavelength, and the OFC spectra were obtained by Fourier transform of the interferograms. The sampling points of the discrete Fourier transform need to precisely match the positions of the OFC modes contained in the interfereogram to surpass the Fourier resolution limit. Therefore, we added corrections to the sampling frequency: correction for offset frequency mismatch caused by the OFC offset frequency, and correction for sampling point spacing mismatch caused by the slight difference between the actual interferogram length and $c/f_{rep}$. Accurate correction of the sampling point spacing mismatch requires an "effective" He-Ne laser wavelength that accounts for the refractive index of air and the effect of laser beam alignments, which is difficult to be accurately estimated. To find an effective He-Ne wavelength we follow the procedure described in detail in ref [2]. In this work, the ILS that occurs due to the mismatch between sampling points and comb lines was evaluated through the residuals of the line-shape fitting of absorption lines. The effective He-Ne wavelength was set to the value that minimized the ILS effect in the low pressure data. The same effective wavelength was used for the high pressure data.

In addition to this procedure, we added a correction for interferogram flatness. In this work, the interferogram length corresponding to $c/f_{rep}$ is relatively long at 2.4 m, and the intensity difference between the center and edge of an interferogram is not negligible. The acquired non-flat interferogram acts in the same way as a window function in FTIR, resulting in ILS effects on the resulting absorption lines. Figures 3(a) and (b) show a reference interferogram obtained with a near-infrared cw laser for the flatness correction and an interferogram of the mid-IR OFC, respectively. The reference interferogram was measured by replacing the input fiber with a near infrared one. To avoid chromatic aberration, a reflective fiber collimator was used for the input part of the FTS system and lenses were removed from the setup. After optimizing the alignment, the intensity difference of the interferogram at the center and at the edge was 5 %. The envelope of the cw laser interferogram was obtained after spline interpolation and used as a correction curve to compensate for the varying amplitude of the OFC interferogram. Figure 3(c) shows the absorption spectra measured at 10 Torr, R7 line with flatness correction (green) and without correction (red). The lower plot shows the effect of the correction as the difference between the two line shapes. The difference was about 0.5 % at the center of the line, causing 4 % deviation in the fitted Lorentzian width. Since the broadening and shift effects caused by beam divergence in FTS are absolute wavelength dependent, intrinsic residual ILS effects due to the amplitude correction using the different wavelength cw laser in general could be expected [38]. On the other hand, we found those effects much smaller than predicted for non-comb FTS, the influence of the amplitude correction in the high-pressure region is very small; the change of the fitted Lorentzian width at 400 Torr due to the interferogram amplitude correction was only 0.007 %. In addition, the expected line asymmetry predicted by the wavenumber-dependence is not observed in fig. 3(c) within the SNR of this experiment.

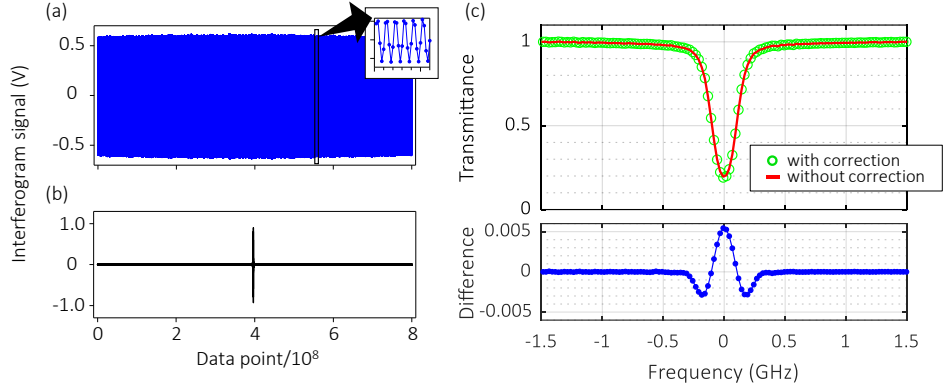

Fig. 3. (a) Interferogram of the near-infrared cw laser and a magnified view. The envelope of the interferogram was used for the flatness correction. (b) Interferogram of the mid-IR OCF. The interferogram length is about 2.4 m, corresponding to $c/f_{rep}$ (c) Comparison of the line shapes with correction (green circle) and without the flatness correction (red line) in the R7 line observed at 10 Torr. The difference between the line shapes is shown in blue in the lower plot.

## 3. Results and discussion

Magnified views of the R7 line measured at each pressure are shown in the top plots of Fig. 4, and the residuals of the Voigt profile (VP), Nelkin-Ghatak profile (NGP), and speed-dependent Voigt profile (SDVP) fits are shown below the plots of the line profiles. The VP is the simplest profile, being a convolution of Gaussian and Lorentzian profiles corresponding to Doppler and collisional broadening and shift, respectively. The NGP [33] is a profile that treats the Dicke narrowing in the hard-collision model without the effect of speed-dependence of collisional parameters. The SDVP [31] is a profile that takes into account the dependence of collision parameters on the emitter velocity but without the Dicke narrowing. In the fitting procedure, we fitted the parameters of a linear baseline, baseline etalons, line center frequency, line intensity, and collisional broadening for all the profiles, the reduced effective frequency of velocity changing collisions for the NGP, and the coefficient of speed-dependent collisional width ($a_w$) with the quadratic approximation for the SDVP [39]. The Doppler broadening width was fixed to the value calculated from the sample cell temperature, which is approximately 150 MHz. For the low-pressure data obtained from interleaved measurements, we fit the baselines individually for each constituent spectrum to avoid SNR degradation caused by phase drifts of the baseline etalons. Fitting the baselines individually improved the quality of the fit (QF) by approximately 100, as compared to fitting the baseline after interleaving. Here, the QF is defined as the ratio of the peak absorption to the standard deviation of the fit residuals.

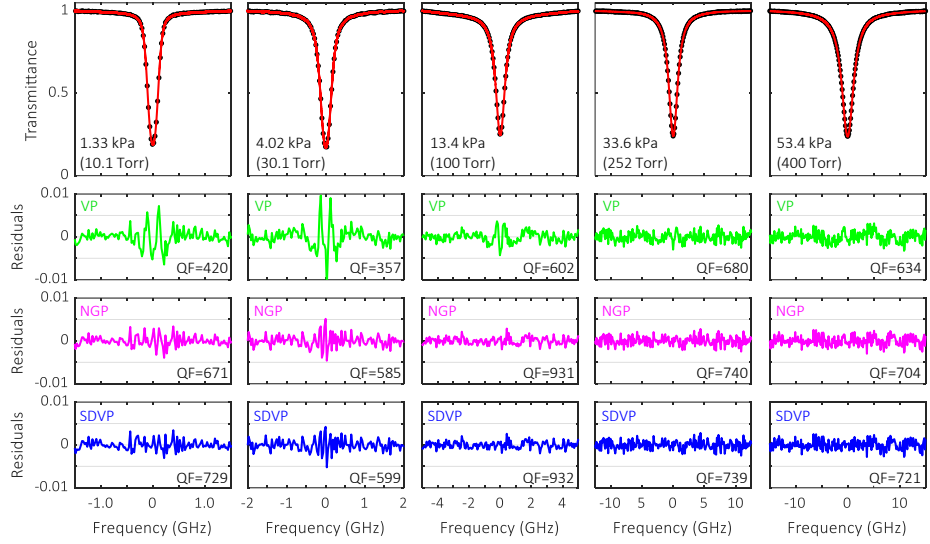

Fig. 4. Observed line shape of R7 line of CO perturbed by $N_2$ at each pressure (black circle) and fitted SDVP line shapes (red). Residuals of VP fittings (green), NGP (magenta), and SDVP (blue). The QFs are shown in the residual plots.

The w-shape systematic deviation from the VP is significant in the pressure range of 10 to 100 Torr, indicating that we have achieved highly precise measurements beyond the accuracy of VP and that the advanced line-shape models are necessary in the analysis. Indeed, NGP and SDVP fits agree better with the experiment than the VP over the entire pressure range and SDVP shows better agreement than NGP at 10 Torr and 30 Torr; compared with the VP fit, the QF of SDVP fits is about 1.7 times larger at 10 and 30 Torr and 1.5 times larger at 100 Torr. From 100 to 400 Torr, there is no significant difference in the QF between NGP and SDVP, with some QFs of the SDVP being slightly better than those of the NGP. Similar remarks regarding QFs of different profile fits and their pressure dependence apply to all the other measured lines.

Previous studies have shown that both Dicke narrowing, and speed-dependent collision effects are significant in the CO-$N_2$ system [40]. However, our attempted single line fits of the SDNGP have failed to converge because of strong correlation between the Dicke narrowing and speed-dependent parameters [39,41]. Lacking theoretical values at which one of these parameters could be fixed, we simply use SDVP, which represents the experimental spectrum the best. Analyses of collisional parameters in this paper in subsequent paragraphs were all based on the results of SDVP fits.

The fitted values of Lorentzian HWHM, $a_w$, and center frequency of the R7 and P7 lines for each pressure are plotted in Fig. 5. The error bars in the Lorentz width plots (Fig. 5(a) and (b)) represent 1σ of the fitting uncertainties of the parameter, which are smaller than the circles in the plots. We used the error bars as weights in linear fits (red), whose slope gives the coefficients of pressure-independent collisional broadening ($\gamma$). The error bars in the $a_w$ plots (Fig. 5(c) and (d)) also represent 1σ of the fitting uncertainties. The $a_w$ of a transition line is calculated from weighted average of $a_w$ at each pressure. The error bars of $a_w$ are relatively large for their absolute values, which is considered to be limited by the SNR of the obtained spectra. If Dicke narrowing was prominent, the value of $a_w$ would appear to be smaller at higher pressures, and it would be possible to fit the Dicke narrowing parameter together in a multi-line fit procedure. However, since no significant variation of $a_w$ with pressure was observed in the present measurements, the obtained data SNR does not seem to allow a determination of the relative contributions of speed-dependent effects and Dicke narrowing. Therefore, the multi-line fit was not applied to our data.

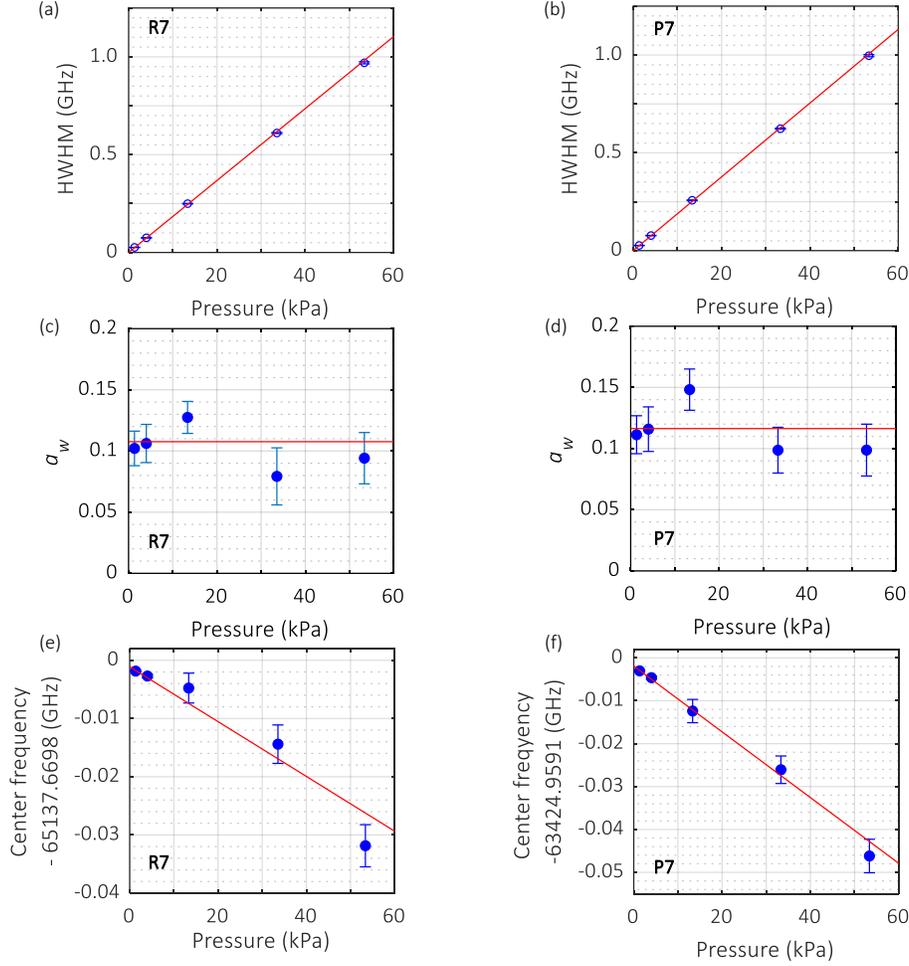

Fig. 5. Fitted Lorentzian HWHM vs. pressure of (a) R7 and (b) P7 lines. The $a_w$ vs. pressure of (c) R7 and (d) P7 lines. Center frequency shift against the pressure of (e) R7 and (f) P7 lines. Linear fits to the data were performed using the error bars as the weight of the fits (red).

The error bars in the center frequency plots (Fig. 5(e) and (f)) are 1σ of the fitting uncertainty for 10 and 30 Torr. The uncertainty of the effective He-Ne wavelength, i.e., the uncertainty of the FTS sampling frequency spacing, is not considered at these pressures because we determined it by minimizing the ILS effects in the observed data. For the higher-pressure data, in addition to the fitting uncertainty, we added 3σ of the FTS sampling point deviation caused by the variation of the effective He-Ne wavelength. The frequency stability of the commercial stabilized He-Ne laser used in our setup is in the $10^{-8}$ level, resulting in an FTS sampling point uncertainty of less than 0.4 MHz in the mid-IR region. However, during our measurement (two days for all pressures), the drift of this He-Ne laser was larger than the specification, and the calculated 3σ value was 3.3 MHz. As with the collisional broadening, the collisional shift coefficients (δ) were obtained by linear fits using the error bars as weights. Here, the speed-dependent collisional shift is negligibly small. The linear fits to the P7 line results in good agreement with the data in the range of the error bars. However, for the R7 line result, the observed data were slightly off from the linear fit. This indicates that the drift of the He-Ne laser wavelength was incidentally larger than 3σ when the high-pressure measurement was performed.

Table 1: Collisional line-shape parameters of CO-$N_2$.

| $m$ | $\gamma$ (cm$^{-1}$/atm) | $a_w$ | $\delta$ (cm$^{-1}$/atm) 10$^{-3}$ | $m$ | $\gamma$ (cm$^{-1}$/atm) | $a_w$ | $\delta$ (cm$^{-1}$/atm) 10$^{-3}$ |
|---|---|---|---|---|---|---|---|
| -1 | 0.0835 (74) | 0.083 (69) | -0.86 (17) | 1 | 0.0817 (42) | 0.096 (08) | -0.89 (54) |
| -2 | 0.0791 (31) | 0.110 (40) | -1.77 (19) | 2 | 0.0765 (20) | 0.086 (19) | -1.24 (27) |
| -3 | 0.0749 (18) | 0.120 (16) | -1.69 (52) | 3 | 0.0737 (30) | 0.108 (12) | -1.58 (37) |
| -4 | -- | -- | -- | 4 | 0.0685 (34) | 0.111 (13) | -1.57 (57) |
| -5 | 0.0672 (30) | 0.114 (19) | -2.26 (28) | 5 | 0.0677 (14) | 0.114 (09) | -1.58 (65) |
| -6 | 0.0649 (22) | 0.103 (17) | -2.58 (49) | 6 | 0.0651 (22) | 0.111 (16) | -1.55 (57) |
| -7 | 0.0637 (22) | 0.116 (21) | -2.59 (39) | 7 | 0.0637 (17) | 0.112 (11) | -1.53 (54) |
| -8 | 0.0627 (18) | 0.113 (17) | -2.73 (41) | 8 | 0.0622 (22) | 0.108 (17) | -1.59 (57) |
| -9 | 0.0615 (18) | 0.110 (12) | -2.71 (34) | 9 | 0.0610 (26) | 0.101 (17) | -1.51 (54) |
| -10 | 0.0615 (16) | 0.115 (14) | -2.62 (09) | 10 | 0.0608 (25) | 0.112 (22) | -1.51 (42) |
| -11 | 0.0601 (12) | 0.106 (04) | -2.80 (30) | 11 | 0.0599 (16) | 0.102 (11) | -1.53 (53) |
| -12 | 0.0597 (24) | 0.117 (22) | -2.83 (35) | 12 | 0.0587 (21) | 0.097 (20) | -1.60 (56) |
| -13 | 0.0588 (25) | 0.115 (20) | -2.87 (33) | 13 | 0.0587 (17) | 0.114 (12) | -1.69 (63) |
| -14 | 0.0579 (18) | 0.109 (22) | -2.85 (21) | 14 | 0.0580 (17) | 0.112 (08) | -1.54 (47) |
| -15 | 0.0566 (33) | 0.125 (35) | -2.79 (23) | 15 | 0.0568 (24) | 0.119 (18) | -1.63 (57) |
| -16 | 0.0566 (31) | 0.131 (21) | -3.00 (31) | 16 | 0.0556 (30) | 0.115 (22) | -1.76 (46) |
| -17 | 0.0565 (43) | 0.149 (54) | -3.12 (37) | 17 | 0.0561 (19) | 0.132 (20) | -1.72 (37) |
| -18 | 0.0545 (41) | 0.140 (49) | -3.00 (34) | 18 | 0.0550 (29) | 0.145 (22) | -1.74 (60) |
| -19 | 0.0515 (39) | 0.159 (48) | -3.30 (63) | 19 | 0.0531 (24) | 0.118 (26) | -1.60 (55) |
| -20 | 0.0522 (37) | 0.145 (72) | -3.00 (40) | 20 | 0.0540 (14) | 0.150 (37) | -2.27 (56) |
| | | | | 21 | 0.0522 (54) | 0.151 (61) | -1.94 (39) |

The collisional broadening, $a_w$, and shift coefficients derived from the weighted linear fits are shown in Table 1 with the line number ($m$), where $m = -J$ for the P branch and $m = J + 1$ for the R branch. This parameters can be transfomed to the form used in HITRAN [26] $\gamma_0 = \gamma$, $\gamma_2 = \gamma a_w$ and $\delta_0 = \delta$. The number in parentheses indicates the uncertainties of the parameters derived as the fit uncertainty in the linear fit. The parameters of the P4 line are not shown in the table because it overlaps with a weak isotopologue line and it could not be fitted separately from the isotopologue line in the high pressure region. Figure 6 (a) and (b) plot the broadening coefficient and $a_w$ as a function of $m$. The error bars show uncertainty of the coefficients. Both the broadening coefficient and $a_w$ exhibit $|m|$ dependence. The $m$ dependence of the broadening coefficient has a characteristic shape that is often seen in other studies.

The broadening coefficients reported in a previous study using cw-laser spectroscopy [23] are shown in Fig. 6(a). Noticeable, systematic offsets are found; this study provides collisional broadening coefficients about 2.3 % larger in average than the previous study. This discrepancy and their sources are discussed in more details below. In Fig. 7, the percentage difference between the broadening coefficients reported in this work ($\gamma$) and in the previous study ($\gamma'$) is plotted in red circles. The uncertainty of our coefficients is shaded on the graph, but many points are outside this range. The broadening factor is temperature dependent and is given by the convenient scaling law $\gamma'/\gamma = (T'/T)^{(-n)}$, where $T'$ and $T$ are the sample gas temperatures at which $\gamma'$ and $\gamma$ are measured at 301 K and 296.5 K, respectively. We have calculated the temperature correction for $\gamma$ using the temperature exponent $n$ reported in ref [23]. The discrepancy of the broadening coefficients at the same temperature is shown in Fig. 7 as green circles. The average difference in coefficients is about 1.3 % even after temperature correction. This residual offset can be attributed to the two different line-shape models used in the fits. Based on the same data set, we compared the broadening coefficients obtained from the VP fit with a temperature correction. The results are plotted as blue circles in Fig. 7, and the average difference is smaller (0.3 %), and all points are within the uncertainty range. It shows that if the same line shape model (in this case VP) was used for analyses of both sets of data, sub-percent agreement was reached. As verified in ref [23], different line-shape models can give slightly different broadening factors. From this result, it is concluded that the difference in broadening coefficient from the previous study seen in Fig. 6 is attributed to the difference in measurement temperature and the fitted line shape models. The influence of residual ILS of our spectrometer is sufficiently small to be negligible with the SNR in the present measurements.

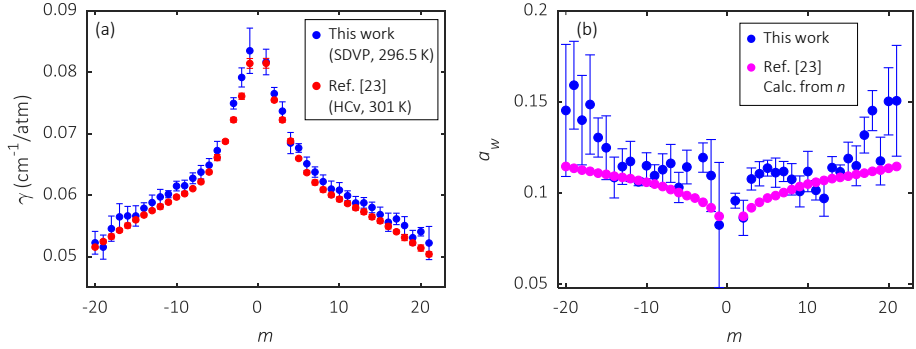

Fig. 6. (a) Broadening coefficients $\gamma$ plotted vs $m$ obtained in this work (blue), reported in Ref. [23] measured with tunable diode laser ($\gamma'$, red). (b) Speed-dependent collisional width ($a_w$) vs $m$. The fitted values to our experimental value (blue), and calculated values from Eq. 1 (magenta).

It has been concluded in the past that the line profile used in the previous study [23], convoluted speed-dependent hard collision (HCv) profile [21], was not adequate to represent speed-dependent collisionally narrowed profiles. The HCv profile is a phenomenological expression proposed as a convolution of weighted sum of speed-dependent Lorentzian profiles [42] and the Dicke-narrowed [32] Doppler distribution affected by hard collisions [33,34]. The profile proposed by Henry et al. [21] was an analog of convolved speed-dependent Galatry profile given by Duggan et al. [39] incorporating soft collisions [43]. Both of these convolved profiles [21,39], had, however, an important drawback, in the limit of no Dicke narrowing they do not reduce to the expected speed-dependent Voigt profile [31]. This problem was solved a few years later in both cases by deriving the speed-dependent Galatry profile [44,45] and the speed-dependent Nelkin-Ghatak profile [29,30] for soft and hard collisions, respectively. Although the SDVP does not include the Dicke narrowing effect, it can well represent Dicke narrowed profiles if narrowing parameter is small comparing to the Doppler width and signal to noise ratio is moderate (order of few hundred) thanks to large correlation between parameters describing Dicke narrowing and speed-dependent effects [40,41]. The data obtained in this study are compatible with the new standard of spectroscopic data in HITRAN.

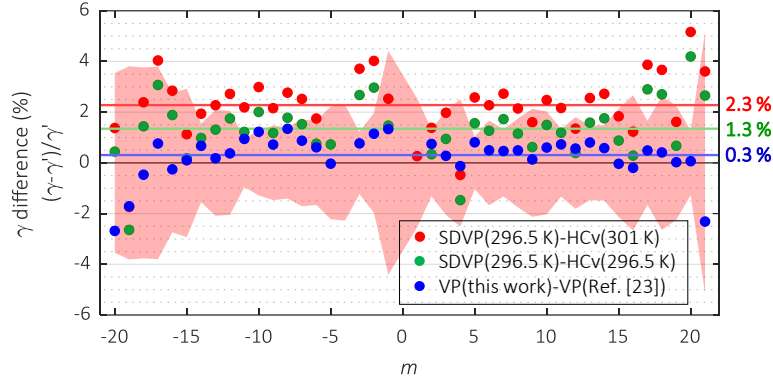

Fig. 7. The difference in broadening coefficients between this study ($\gamma$) derived from SDVP fitting and the previous study ($\gamma'$) [23] using the HCv profile (red). After temperature correction of the previous study data to 296.5 K (green). Comparison between coefficients based on VP fitting at the same temperature (blue). The red shaded area shows the uncertainty of our coefficients. The mean values of each plot are also shown.

On Fig. 6(b) one can see the value of $a_w$ tends to be larger for transitions with large absolute values of $m$. This behavior can be caused by the fact that in our analysis $a_w$ represents all narrowing effects (speed-dependent effects and Dicke narrowing) [40,41]. Therefore, lines with small $|m|$ where correlation between velocity-changing and dephasing or state-changing collisions [34] are expected to reduce Dicke narrowing significantly [46] will lead to lower contribution to values of $a_w$ obtained from the fit than in case of high $|m|$ where Dicke narrowing is expected to be noticeable [46-48]. It will be also affected by line-by-line variations of speed-dependent effects. The connection of temperature dependence of collisional broadening and its speed dependence was recognized by Lance et al. [49] and Brault et al. [50]. Rohart et al. [51] proposed and their followers [52-54] tested estimation of $a_w$ assuming interaction potential having a form of inverse power law. We use here a direct connection of speed-dependence parameter $a_w$ and temperature dependence of collisional broadening derived by Lisak et al. [55]. It allows us to estimate variation of $a_w$ on $|m|$. For this purpose, we use relation derived in [55,56]

$$a_w = \frac{2}{3}\frac{\alpha}{1+\alpha}(1-n) \qquad (1)$$

between the speed dependence parameter $a_w$ and the temperature exponent $n$ at a given perturber/absorber mass ratio $\alpha$. It can be noted that this equation provides similar results to empirical expression obtained for $O_3$ perturbed by $O_2$ or $N_2$ in paper by Tran at al. [57] as well as Eq. (1) can be found in [58]. In the case of $CO$-$N_2$ system investigated here $\alpha = 1$ and above equation takes a very simple form $a_w = (1-n)/3$ [56]. Figure 6(b) shows that $a_w$ estimated in this way using the temperature exponent $n$ reported in Ref. [23] agree well with values obtained from the line shape fitting. Slightly higher fitted values of $a_w$ comparing to estimated ones can be manifestation of Dicke narrowing which was not extracted in our data analysis because insufficient SNR.

The collisional shift coefficients obtained in this study are plotted as blue circles against $m$ in Fig. 8. The error bars show the uncertainty of the coefficients. The shift coefficients reported in Ref. [24] using cw-laser spectroscopy are shown as red circles, and another previous study performed using FTIR [20] is shown as green circles. Our results are close to the results reported in Ref. [24]. The difference between two previous studies or between this study and the FTIR results can be attributed to the errors in the FTIR spectrometer. Due to FTIR phase errors, the quality of line shift measurements by FTIR has long been questioned [59]. For the R-branch coefficients, the results of this study and those of Ref. [24] are in excellent agreement, except for lines with poor SNR. Most of the P branch coefficients also overlap in the area of the error bars.

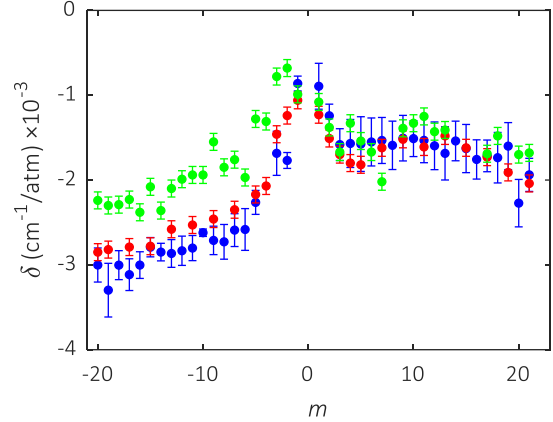

Fig. 8. Shift coefficients plotted vs *m* obtained in this work (blue), reported in Ref. [24] measured with cw-laser difference frequency generation spectrometer (red), and in Ref. [20] measured with FTIR (green).

## 4. Summary

In this work, a precise and accurate line-shape study of the fundamental vibrational band of CO-$N_2$ was performed using a mid-IR OFC-FTS spectrometer. The spectrometer successfully provided accurate line shapes of a large number of lines under several pressure conditions in a short time, 11 s for single acquisition and 9 minutes for 50 averages. All the data shown in this paper were obtained in two working days including the time for gas exchange and maintenance of the mid-infrared frequency comb. The accuracy of the spectral data obtained by conventional FTIR was less reliable than that of the tunable cw laser-based measurements due to phase errors and the effect of ILS. On the other hand, the results of this study demonstrated very good agreement with previous studies using the difference frequency generation technique [23,24], which has been the gold standard for molecular spectroscopy in this spectral range. As shown in Fig. 7, the difference between the broadening coefficients obtained from the VP fit is on average 0.3% for the 41 transitions, and all values agree within the uncertainty of the fit. The shift coefficients, which are difficult to precisely determine due to the small value in CO-$N_2$ system, also show good agreement for most of the lines, as shown in Fig. 8. We used more appropriate line profiles than in previous studies, which are compatible with the new generation spectroscopic data base HITRAN and provided coefficients of pressure broadening and shift for 41 lines, as well as parameters describing speed dependence of collisional width that have not been reported in previous studies. Corrections to the broadening coefficient reported by [23] were on average about 2%.

The spectral data of CO is of high importance to the spectroscopic community due to the role of atmospheric CO in the carbon cycle and its versatility as a probe in various astronomical observations. Consistent sets of spectroscopic data going beyond Voigt profile and valid in wide range of condition are crucial for improvement of gas sensors [60]. Obtaining spectral data from a wider temperature range and comparing it with theoretical calculations would be a useful direction for future studies. OFC-FTS is a powerful spectroscopic tool for providing line-shape parameters to spectral databases, because of its great advantage of simultaneously acquiring high-precision spectra over a broadband measurement range. The spectral data provided by OFC-FTS will contribute to the development of atmospheric science and astronomical observations.

**Funding**

This research was supported by National Science Centre, Poland project no. 2019/35/D/ST2/04114. GK acknowledges the support from the European Union's Horizon 2020 Research and Innovation Program under Marie Sklodowska-Curie Grant Agreement No 101028278.

**Acknowledgement**

We thank Dr hab. inż. Grzegorz Soboń (Wroclaw University of Science and Technology) for useful discussions on the development of the mid-IR comb system and for providing the PCF fiber for the OPO setup.